\documentclass[journal ]{new-aiaa}
\usepackage[utf8]{inputenc}
\usepackage{textcomp}

\usepackage{graphicx}
\usepackage{amsmath}

\usepackage[version=4]{mhchem}
\usepackage{siunitx}
\usepackage{longtable,tabularx}
\usepackage{algorithm}
\usepackage{algpseudocode}
\usepackage{rotating} 

\title{A Methodology to Identify Physical or Computational Experiment Conditions for Uncertainty Mitigation}

\author{Efe Y. Yarbasi\footnote{Graduate Research Associate, Aerospace Systems Design Laboratory, Daniel Guggenheim School of Aerospace Engineering, AIAA Student Member} and Dimitri Mavris\footnote{S.P. Langley Distinguished Regents Professor, Aerospace Systems Design Laboratory, Daniel Guggenheim School of Aerospace Engineering, AIAA Fellow}}
\affil{Georgia Institute of Technology, Atlanta, GA, 30332, United States}

\begin{document}

\maketitle

\begin{abstract}
Complex engineering systems require integration of simulation of sub-systems and calculation of metrics to drive design decisions. This paper introduces a methodology for designing computational or physical experiments for system-level uncertainty mitigation purposes. The methodology follows a previously determined problem ontology, where physical, functional and modeling architectures are decided upon. By carrying out sensitivity analysis techniques utilizing system-level tools, critical epistemic uncertainties can be identified. Afterwards, a framework is introduced to design specific computational and physical experimentation for generating new knowledge about parameters, and for uncertainty mitigation.  The methodology is demonstrated through a case study on an early-stage design Blended-Wing-Body (BWB) aircraft concept, showcasing how aerostructures analyses can be leveraged for mitigating system-level uncertainty, by computer experiments or guiding physical experimentation. The proposed methodology is versatile enough to tackle uncertainty management across various design challenges, highlighting the potential for more risk-informed design processes.
\end{abstract}

\section*{Nomenclature}

\noindent

{\renewcommand\arraystretch{1.0}
\noindent\begin{longtable*}{@{}l @{\quad=\quad} l@{}}
$\alpha$ & Angle of attack, $deg$ \\ 
$C_L$ & Lift coefficient \\
$C_D$ & Drag coefficient    \\
$n$ & Scale of the subscale model \\
$E$ & Young's Modulus, $GPa$ \\
$EI$ & Bending Stiffness, $Nm^2$ \\
$GJ$ & Torsional Stiffness, $Nm^2$ \\
$h$ & Altitude,  $m$ \\
$Re$ & Reynolds number \\
$Ma$ & Mach number \\
$\rho$ & Density, $kg/m^3$ \\
$L$ & Reference length, $m$ \\
$S_T^i$ &Total Sobol Sensitivity Index \\
$\__F$ & Quantity pertaining to the full scale model \\
$\__S$ & Quantity pertaining to the subscale model \\
\end{longtable*}}

\section{Introduction and Background}

The design of a flight vehicle is a lengthy, expensive process spanning many years. With the advance in computational capabilities, designers have been relying on computer models to make predictions about the real-life performance of an aircraft. However, the results obtained from computational tools are never exact due to a lack of understanding of physical phenomena, inadequate modeling and abstractions in product details~\cite{Thunnissen2003,Oberkampf2010,Nolan2015}. The vagueness in quantities of interest is called \textit{uncertainty}. The uncertainty in simulations may lead to erroneous predictions regarding the product; creating \textit{risk}. 

Because most of the cost is committed early in the design~\cite{Mavris2000a}, any decision made on quantities involving significant uncertainty may result in budget overruns, schedule delays and performance shortcomings, as well as safety concerns. Reducing the uncertainty in simulations earlier in the design process will reduce the risk in the final product. The goal of this paper is to present a systematic methodology to identify and mitigate the sources uncertainty in complex, multi-disciplinary problems such as aircraft design, with a focus on uncertainties due to a lack of knowledge (\textit{i.e.}, \textit{epistemic}).



\subsection{The Role of Simulations in Design}

\begin{figure}[h]
	\centering
	\includegraphics[width=0.8 \linewidth]{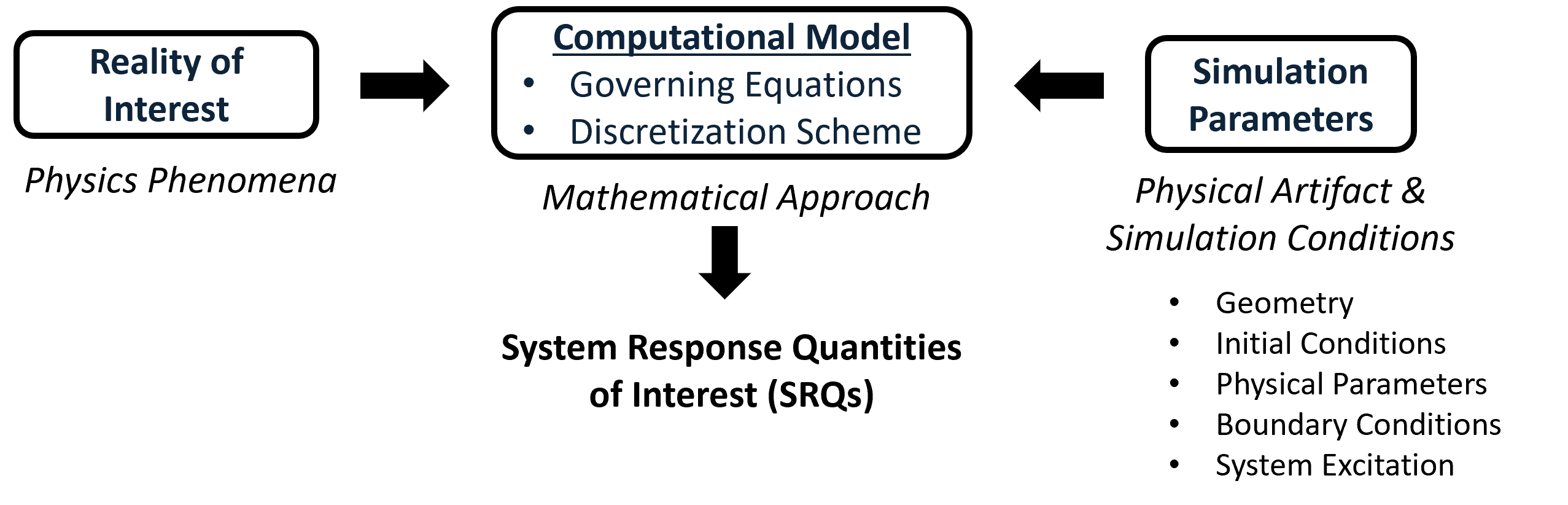}
	\caption{Illustration of a generic computer simulation and how System Response Quantities are obtained. Adapted from \cite{Roy2010}}
	\label{fig:computational_model}
\end{figure}

Computational tools are almost exclusively used to make predictions about the response of a system under a set of inputs and boundary conditions~\cite{Oberkampf2004a}. At the core of computational tools lies a model, representing the reality of interest, commonly in the form of mathematical equations that are obtained from theory or previously measured data. How a computer simulation represents a reality of interest is summarized in Figure~\ref{fig:computational_model}.  Development of the mathematical model implies that there exist some information about the reality of interest (\textit{i.e.}, a physics phenomenon) at different conditions so that the form of the mathematical equation and the parameters that have an impact on the results of the equation can be derived. The parameters include the coefficients and mathematical operations in the equations, as well as anything related to representing the physical artifact, boundary and initial conditions, system excitation~\cite{Roy2010}. A complete set of equations and parameters are used to calculate the system response quantities (SRQ). Depending on the nature of the problem, the calculation can be straightforward or may require the use of some kind of discretization scheme.

If the physics phenomenon is understood well enough such that the form of the mathematical representation is trusted, a new set parameters in the equations (e.g., coefficients) may be sought in order to better match the results with an experimental observation. This process is called \textit{calibration}. With the availability of data on similar artifacts, in similar experimental conditions; calibration enables the utilization of existing simulations to make more accurate predictions with respect to the measured "truth`` model. 

Most models are \textit{abstractions} of the reality of interest as they do not consider the problem at hand in its entirety but only in general aspects that yield useful conclusions without spending time on details that do not significantly impact the information gain. Generally, models that abstract fewer details of the problem are able to provide more detailed information with a better accuracy, but they require detailed information about the system of interest and external conditions to work with. Such models are called \textit{high-fidelity models}. Conversely, \textit{low-fidelity models} abstract larger chunks of the problem and have a quick turnaround time. They may be able to provide valuable information without so much effort going into setting up the model; unlike high-fidelity models, they can generally do it with very low computational cost. The choice of the fidelity of the model is typically left to the practitioner and depends on the application. 

A few decades ago, engineers \textit{had to} rely on physical experiments to come up with new designs or tune them as their computational capabilities were insufficient. Such experiments that include artifacts and instrumentation systems are generally time consuming and expensive to develop. In the context of air vehicle design, design is an inherently iterative process and these experiments would need to be rebuilt and reevaluated. Therefore, while they are effective for evaluation purposes, they cannot be treated as parametric design models unless they have been created with a level of adjustability. With the advance of more powerful computers and widespread use of them in all industries, engineers turned to simulations to generate information about the product they are working on. Although detailed simulations can be prohibitively expensive in terms of work-hours and computation time; the use of computer simulations is typically cheaper and faster than following a build-and-break approach for most large-scale systems.

As the predictions obtained from simulations played a larger part in the design, concepts such as ``\textit{simulation driven design}" has been more prominent in many disciplines.~\cite{Karlberg2013} If the physics models are accurate, constructing solution environments with very fine grids to capture complex physics phenomena accurately become possible. The cost of making a change in the design increases exponentially from initiation to entry into service~\cite{Schrage1995}. If modeling and simulation environments that accurately capture the physics are used in the design loop, it will be possible to identify necessary changes earlier. Because making design changes later may require additional changes in other connected sub-systems, it will lead to an increase in the overall cost~\cite{Mavris1998}.

\subsection{Modeling Physical Phenomena}
 

When the task of designing complex products involves the design of certain systems that are unlike their counterparts or predecessors, the capability of known physics-based modeling techniques may come short. For example, the goal of making predictions about a novel aircraft configuration aircraft, a gap is to be expected between the simulation predictions and the measurements from the finalized design. If the tools are developed for traditional aircraft concepts (e.g., tube and wing configurations) there might even be a physics phenomenon occurring that will not be expected or captured. Even if there is none, the accuracy of models in such cases are still to be questioned. There are inherent abstractions pertaining to the design and the best way to quantify the impact of variations in the quantities of interest by changing the geometric or material properties is by making a comparative assessment with respect to the historical data. However, in this case, historical data simply \textit{do not exist}. 


Because of a lack of knowledge or inherent randomness, the parameters used in modeling equations, boundary/initial conditions, and the geometry are inexact, \textit{i.e., uncertain}. The uncertainty in these parameters and the model itself, manifest themselves as uncertainty in the model output. As mentioned before, any decision made on uncertain predictions will create risk in the design. In order to tackle the overall uncertainty, the sources of individual uncertainties must be meticulously tracked and their impact on the SRQs need to be quantified. By studying the source and nature of these constituents, they can be characterized and the necessary next steps to reduce them can be identified.

In a modeling and simulation environment, every source of uncertainty has a varying degree of impact on the overall uncertainty.  Then, they can be addressed by a specific way depending on their nature. If they are present because of a lack of knowledge about it (\textit{i.e., epistemic uncertainty}), can by definition be reduced~\cite{Caicedo2016}. The means to achieve this goal can be through designing a study or experiment that would generate new information about the model or the parameters in question. In this paper, the focus will be on how to design a \textit{targeted experiment} for uncertainty reduction purposes. Such experiments are not be a replication of the same experimental setup in a more \textit{trusted} domain, but a new setup that is tailored specifically for generating new knowledge pertaining to that source of uncertainty. 


An important consideration is in pursuing targeted experiments is the allowed time and budget of the program. If a lower-level, targeted experiment to reduce uncertainty is too costly or carries even more inherent unknowns due to its experimental setup, it might be undesirable to pursue by the designers. Therefore, these lower-level experiments must be analyzed on a case-by-case basis, and the viability need to be assessed. There will be a trade-off on how much reduction in uncertainty can be expected, against the cost of designing and conducting a tailored experiment. From a realistic perspective, only a limited number of them can be pursued. Considering the number of simulations used in the process of design of a new aircraft, trying to validate the accuracy of every parameter or assumption of every tool will lead to an insurmountable number of experiments. For the ultimate goal of reducing the overall uncertainty, the sources of uncertainty that have the greatest impact on the quantities of interest must be identified. Some parameters that have relatively low uncertainty may have a great effect on a response whereas another parameter with great uncertainty may have little to no effect on the response.

\begin{figure}[]
	\centering
	\includegraphics[width=0.7\linewidth]{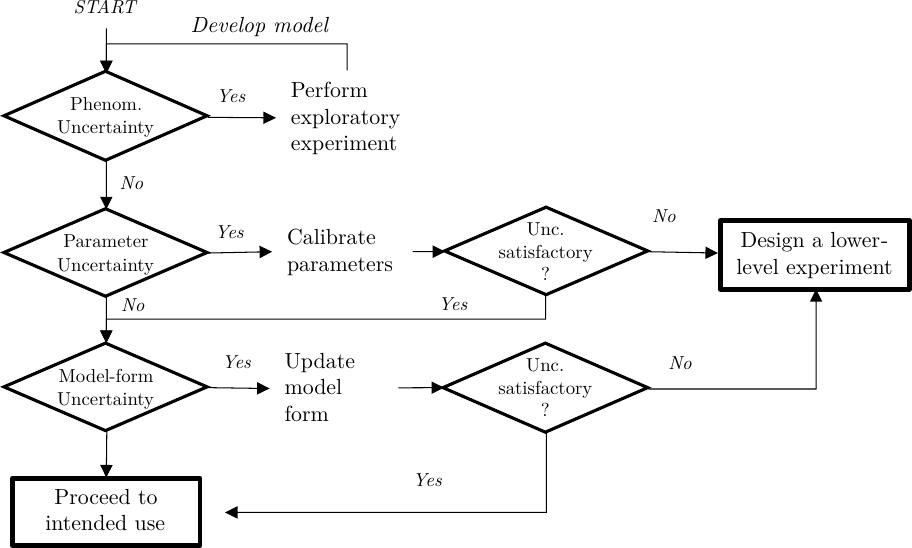}
	\caption{Framework for uncertainty identification and reduction.}
	\label{fig:uncertainty_framework}
\end{figure}

In summary, the train of thought that leads to experimentation to reduce the epistemic uncertainty in modeling and simulation environments is illustrated in Figure~\ref{fig:uncertainty_framework}. If the physics of the problem are relatively well understood, then a computational model can be developed. If not, one needs to perform \textit{discovery experiments}, simply to learn about the physics phenomenon~\cite{Oberkampf2001}. Then, if the results of this model are consistent and accurate, then it can be applied to the desired problem. If not, aforementioned lower-level experiments can be pursued to reduce the uncertainty in the models. Created knowledge should enable the reduction of uncertainty in the parameters, or the models; reducing the overall uncertainty.

\subsection{Reduced-Scale Experimentation}
\label{sec:reduced_experimentation}

An ideal, accurate physical test would normally require the building duplicates of the system of interest in the corresponding domain so that obtained results reflect actual conditions. Although this poses little to no issues for computational experiments -barring inherent modeling assumptions-, as the scale of the simulation artifact does not matter for a computational tool, it has major implications on the design of ground and flight tests. Producing a full-scale replica of the actual product with its all required details for a certain simulation is a expensive and difficult in the aerospace industry. Although full-scale testing is necessary for some cases and reliability/certification tests~\cite{Simitses2001}, it is desirable to reduce the number of them. Therefore, engineers always tried to duplicate the full-scale test conditions on a reduced-scale model of the test artifact.


The principles of \textit{similitude} have been laid out by Buckingham in early 20th century~\cite{Buckingham1914,Buckingham1915}. By expanding on Rayleigh's method of dimensional analysis\cite {LordRAYLEIGH1915,Buckingham2009}, he proposed the \textit{Buckingham-Pi Theorem}. This method enables to express a complex physical equation in terms of dimensionless and independent quantities ($\Pi$ groups). Although they need not to be unique, they are independent and form a complete set. These non-dimensional quantities are used to establish similitude between models of different scales. When a reduced scale model satisfies certain \textit{similitude conditions} with the full-scale model, the same response is expected between the two. Similitudes can be categorized in three different groups\cite{Casaburo2019}:

\begin{enumerate}
	\item \textit{Geometric similitude:} Geometry is equally scaled  
	\item \textit{Kinematic similitude:} ``Homologous particles lie at homologous points at homologous times"~\cite{Baker1991}
	\item \textit{Dynamic similitude:} homologous forces act on homologous parts or points of the system. 
\end{enumerate}

\subsection{Identification of Critical Uncertainties via Sensitivity Analysis} 

Over the recent decades, the variety of problems of interest has led to the development of many sensitivity analysis techniques. While some of them are \textit{quantitative} and \textit{model-free}~\cite{Saltelli2002}, some depend on the specific type of the mathematical model used~\cite{Borgonovo2016}. Similar to how engineering problems can be addressed with many different mathematical approaches,  sensitivity analyses can be carried out in different ways. For the most basic and low-dimensional cases, even graphical representations such as scatter plots may yield useful information about the sensitivities~\cite{Saltelli2008}. As the system gets more complicated however, methods such as local sensitivity analyses (LSA), global sensitivity analyses (GSA) and regression-based tools such as prediction profilers may be used.

LSA methods provide a local assessment of the sensitivity of the outputs to changes in the inputs, and are only valid near the current operating conditions of the system. GSA methods are designed to address the limitations of LSA methods by providing a more comprehensive assessment of the sensitivity of the outputs to changes in the inputs. Generally speaking, GSA methods take into account the behavior of the system over a large range of input values, and provide a quantitative measure of the relative importance of different inputs in the system. In addition, GSA methods do not require assumptions about the relationship between the inputs and outputs, such as the function being differentiable, and are well suited for high-dimensional problems. Therefore, GSA methods has been dubbed the \textit{golden standard} in sensitivity analysis, in the presence of uncertainty~\cite{Saltelli2005}.


Variance-based methods apportion the variance caused in the model outputs with respect to the model input and their interactions. One of the most popular variance-based methods is Sobol Method~\cite{Sobol2001}. Consider a function $f(X) = Y$, Sobol index for a variable $X^i$ is the ratio of the variability obtained by calculating variability due to all other inputs to the overall variability in the output. These variations can be obtained from parametric or non-parametric sampling techniques. Following this definition, the first-order effect index of input $X^i$ can be defined as: 

\begin{equation}
	S^{i}_1 = \frac{V_{X^i}(E_{X^{-i}}(Y\mid X^i))}{V(Y)}
\label{eq:sobol_first}
\end{equation}

where the denominator represents the total variability in the response $Y$ whereas the numerator represents the variation of $Y$ while changing $X^i$ but keeping all the other variables constant. The first-order effect represents the variability caused by $X^i$ only. Following the same logic, combined effect of two variables $X^i$ and $X^j$ can be calculated:

\begin{equation}
	S^{i}_1 + S^{j}_1 + S^{ij}_2=\frac{V_{X^{ij}}(E_{X^{-ij}}(Y\mid X^{ij}))}{V(Y)}
 \label{eq:sobol_second}
\end{equation}

Finally, since the total effect of a variable will be the sum of all first-order effects and the sum of all interactions of all orders with other input variables. Because the sum of all sensitivity indices must be unity, then the total effect index of $X^i$ can be calculated as~\cite{Homma1996}:

\begin{equation}
 S^{i}_T = 1 - \frac{V_{X^{-i}}(E_{X^{i}}(Y\mid X^{-i}))}{V(Y)}
 \label{eq:sobol_total}
\end{equation}
%
%

Because Sobol method is tool agnostic and can be used without any approximations of the objective function, it will be employed throughout this paper for sensitivity analysis purposes. 

\section{Development of the Methodology}

\begin{figure}[]
	\centering
	\includegraphics[width=0.8\linewidth]{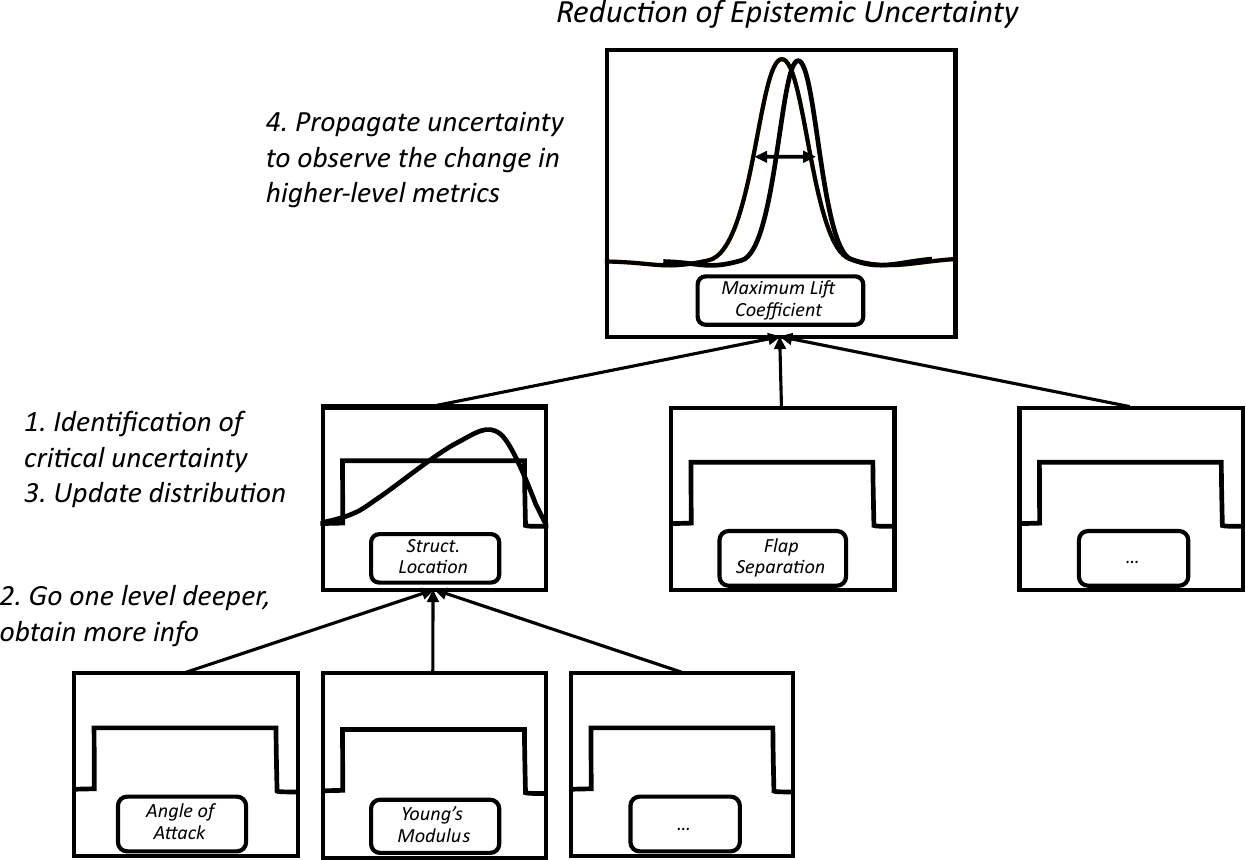}
	\caption{Schematic illustration of the reduction of epistemic uncertainty via following the proposed framework.}
	\label{fig:framework_distributions}
\end{figure}

The proposed methodology is developed with the purpose of identifying and reducing the sources of epistemic uncertainty in complex design projects with a systematic fashion. First, the problem for which the mitigation of uncertainty is the objective, is defined, and corresponding high-level requirements are identified. In this step, the disciplines that are involved in the problem at hand and how requirements flow down to analyses are noted. Then, the problem ontology is formulated by using functional, physical and modeling decompositions. A top-down decision making framework is followed to create candidate, fit-for-purpose modeling and simulation environments and select the most appropriate one for the problem at hand. Upon completion of this step, a modeling and simulation environment that covers every \textit{important} aspect of the problem and satisfies the set of modeling requirements will be obtained, while acknowledging the abstractions of the environment. The third step is to run the model to collect data. Due to aforementioned reasons, there will be many uncertainties in the model. Therefore the focus of on this step is to identify critical uncertainties that have a significant impact on the model response. If one deems this uncertainty to be unacceptable, or to have a significant impact on the decision making processes; then a lower-level experiment will be designed in order to create new knowledge pertaining to that uncertainty. This new knowledge can be carried over to the modeling and simulation environment so that the impact of new uncertainty characteristics (i.e., probability distribution) on the model response can be observed. The main idea of the proposed methodology is to generate new knowledge in a systematic way, and update appropriate components involved in the modeling and simulation environment with the newly obtained piece of information. This process is illustrated in a schematic way in Figure~\ref{fig:framework_distributions}.

\begin{enumerate}
    \item \textbf{Problem Definition:} \\
    All aircraft are designed to answer a specific set of requirements that are borne out of needs of the market or stakeholders. Definition of the \textit{concept of operations} outline the broad range of missions and operations that this aircraft will be used for. After the overall purpose of the aircraft is determined, the designers can decide on the required capabilities and the metrics that these capabilities are going to be measured by can be derived. Considering these information, a concept of the aircraft and the rough size of the aircraft can be determined. This process can be completed by \textit{decomposing the requirements} and tracking their impact on metrics of capability and operations perspectives~\cite{Taylor2023}. Multi-role aircraft will require additional capabilities will emerge and parallel decompositions may need to be used. 

    \item \textbf{Formulate the Problem Ontology:} \\        
    Following systems-engineering based methods given in References~\cite{Taylor2023,Yarbasi2023}, a requirements analysis and breakdown can be performed, identifying the key requirements for the aircraft, the mission and the wing structure to be analyzed. Then, a physical decomposition of the sub-assembly is developed, outlining its key components and their functionalities, decide on the set of abstractions. Finally, a modeling architecture that maps the physical and functional components to the decompositions is created. This mapping from requirements all the way to the modeling and simulation attributes is called the \textit{problem ontology}, and illustrated in Figure~\ref{fig:ontology}. With this step it is possible to follow a decision-making framework to select the appropriate modeling and simulation tool among many candidates for this task. 

    \begin{figure}[h]
	\centering
	\includegraphics[width=0.5\linewidth]{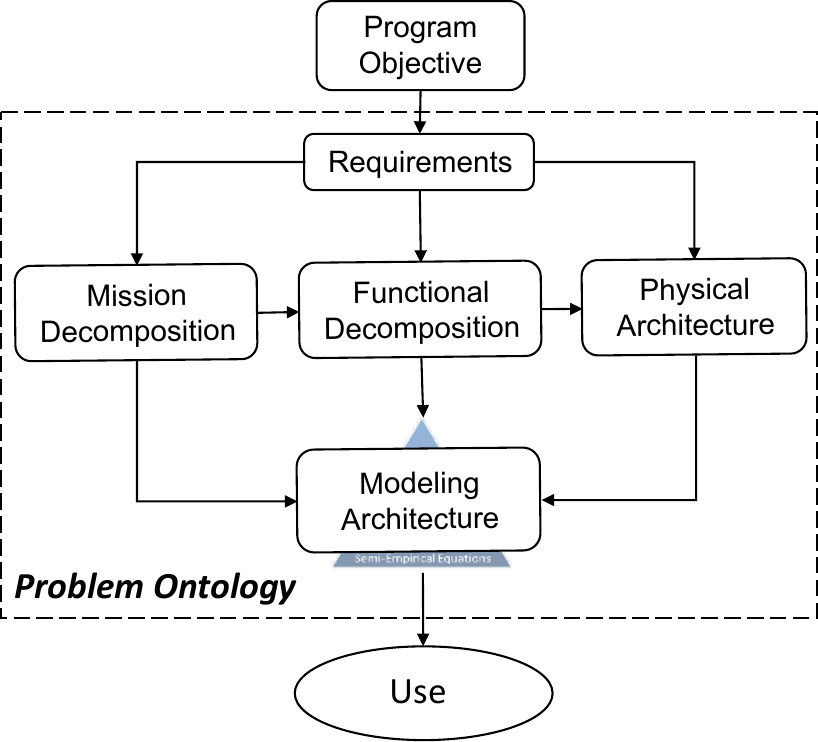}
	\caption{Method used for system decomposition in this work. The construct defined as the \textit{problem ontology} is depicted by the dashed rectangle.}
	\label{fig:ontology}
\end{figure}

    \item \textbf{Identification of Critical Uncertainties} \\
    With a defined M\&S environment, it is possible to run cases and identify the critical uncertainties rigorously that have the most impact on the quantities of interest. As mentioned before, there is a plethora of methods how uncertainty can be mathematically represented and the impact is quantified. For this use case, a sensitivity analysis will be performed to determine which parameters have the greatest impact on the output uncertainty in corresponding tools, by calculating total Sobol indices.

    \item \textbf{Design a Lower-level Experiment:} 

    The next task is to address the identified critical uncertainties at the selected M\&S environment. To this end, the steps corresponding to designing a lower-level experiment are illustrated in Figure~\ref{fig:rq3_overview}. It is essential to note here that the primary purpose of this lower-level experiment is \textit{not} to model the same phenomenon at a subsequent higher-fidelity tool, but rather use the extra fidelity to mitigate uncertainty for system-level investigation purposes. 


    The experimental design is bifurcated into computational experiments (CX) and physical experiments (PX), each may serve a unique purpose within the research context. For computational experiments, the focus is on leveraging computational models to simulate scenarios under various conditions and parameters, allowing for a broad exploration of the problem space without the constraints of physical implementation. Conversely, the physical experiments involve the design and execution of experiments in a physical environment. This phase is intricately linked to the computational experiment by the fact that they can accurately represent the physical experiments can be used to guide the physical experimentation setups. This entails a careful calibration process, ensuring that the computational models reflect the real-world constraints and variables encountered in the physical domain as best as possible. This step is a standalone research area by itself, and it will only be demonstrated on a singular case.
    
    Upon the completion of experimentation procedure, the execution phase takes place, where the experiments are conducted according to the predefined designs. This stage is critical for gathering empirical data and insights, which are then subjected to rigorous statistical analysis. The interpretation of these results forms the basis for drawing meaningful conclusions, ultimately contributing to the generation of new knowledge pertaining to the epistemic uncertainty in question.This methodological approach, characterized by its dual emphasis on computational and physical experimentation, provides a robust framework for analyzing uncertainties.
    
    
\begin{figure}[]
	\centering
	\includegraphics[width=0.9\linewidth]{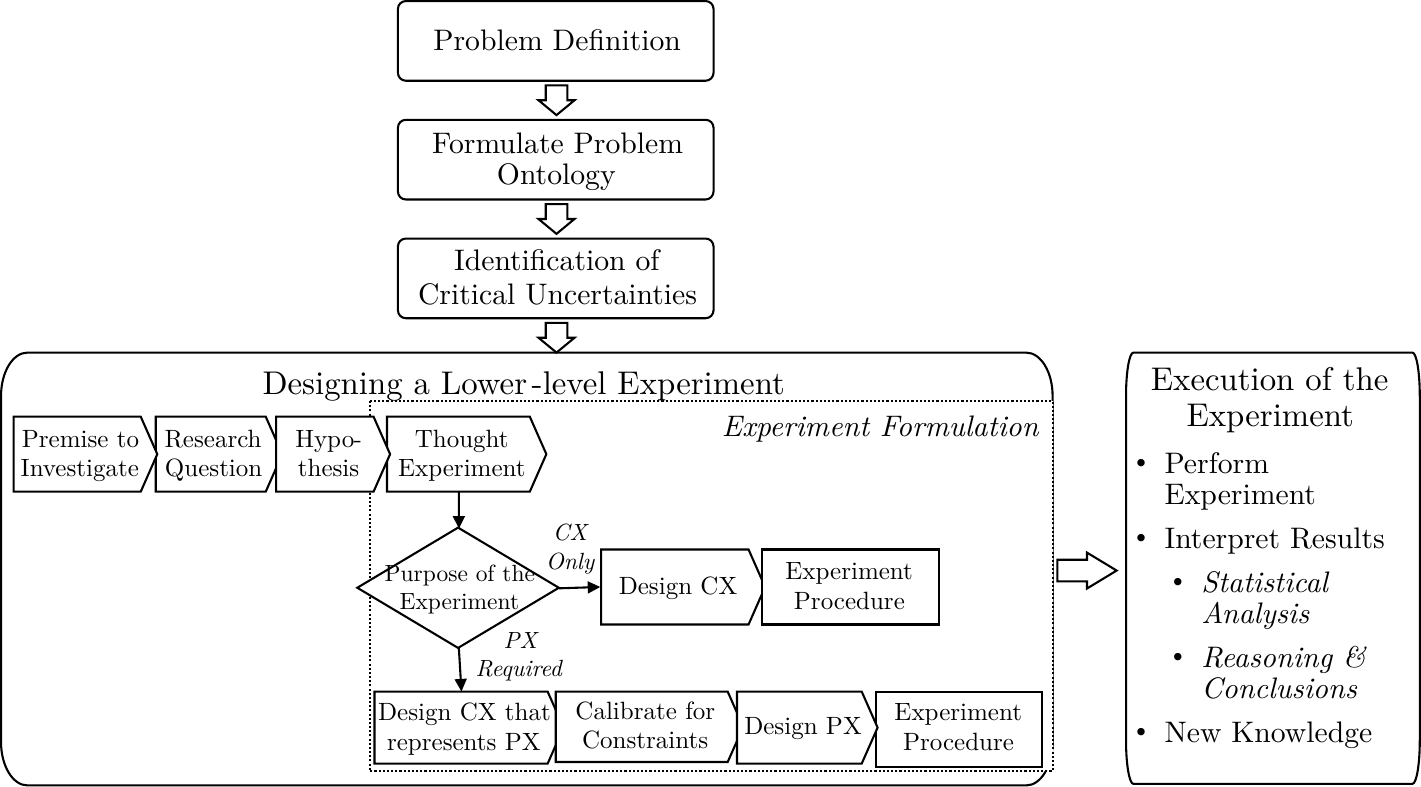}
	\caption{Overview of the methodology followed in this chapter. (PX: physical experiment, CX: computational experiment)}
	\label{fig:rq3_overview}
\end{figure}

    
\end{enumerate}

\section{Demonstration and Results}

\subsection{Formulating the Problem Ontology}

\begin{figure}[]
	\centering
	\includegraphics[width=0.6\linewidth]{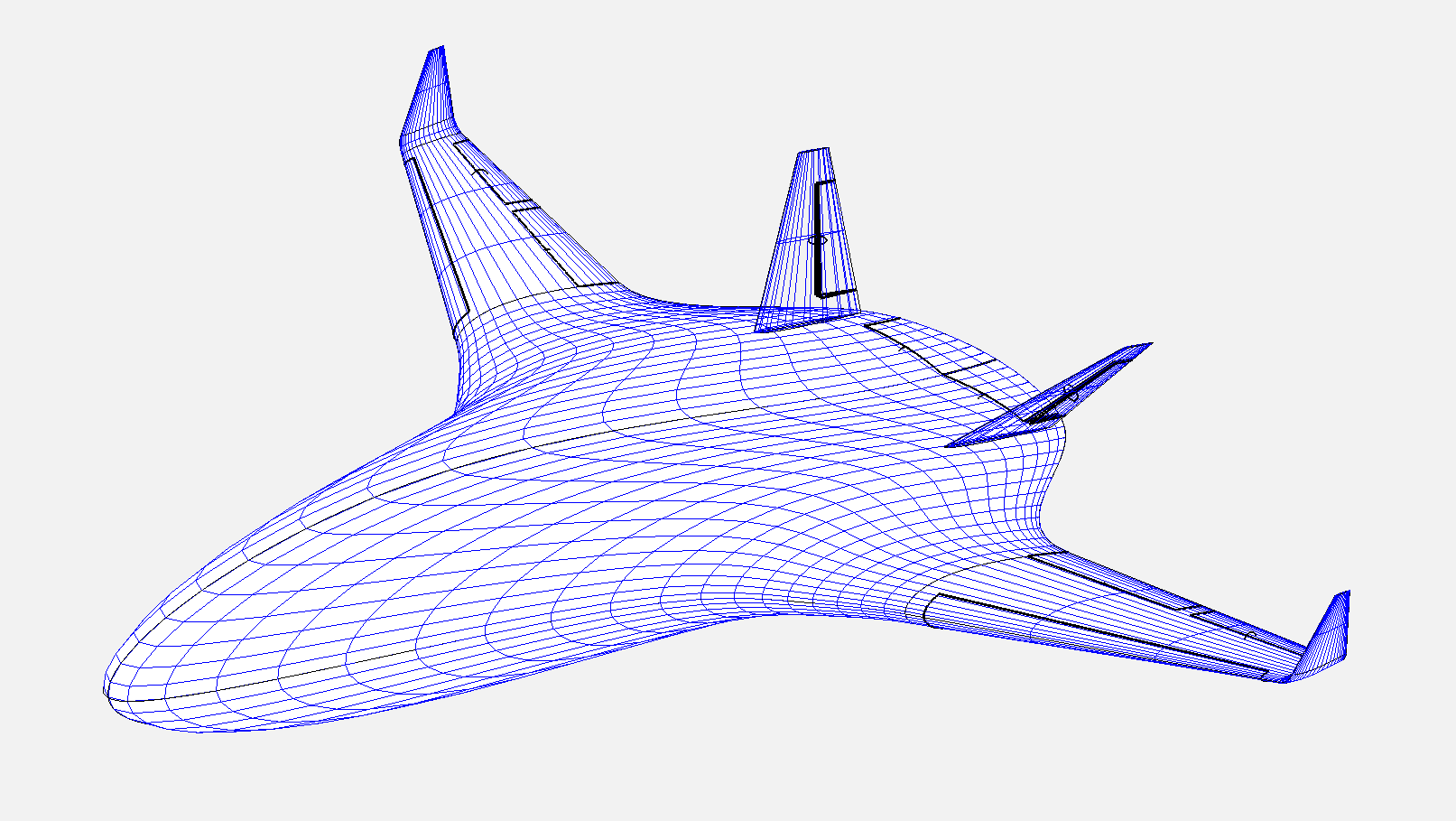}
	\caption{BWB concept used in this work.}
	\label{fig:BWB_VSP}
\end{figure}

Development of a next-generation air vehicle platform involves significant uncertainties. To demonstrate how the methodology would apply on such a scenario, the problem selected is aerostructures analyses for a Blended-Wing-Body (BWB)  aircraft in the conceptual design stage. The goal is to increase confidence in the predictions of the aircraft range by reducing the associated uncertainty on parameters used in design. A representative OpenVSP drawing of the BWB aircraft used in this work is given in Figure~\ref{fig:BWB_VSP}.

\subsection{Identification of Critical Uncertainties}


For the given use case, two tools are found to be appropriate for the early-stage design exploration purposes, FLOPS and OpenAeroStruct.


As a low fidelity tool, NASA's Flight Optimization System (FLOPS)\cite{FLOPS} will be used to calculate the range of the BWB concept for different designs. FLOPS is employed due to its efficiency in early-stage design assessment, providing a quick and broad analysis under varying conditions with minimal computational resources. FLOPS facilitates the exploration of a wide range of design spaces by rapidly estimating performance metrics, which is crucial during the conceptual design phase where multiple design iterations are evaluated for feasibility and performance optimization. FLOPS uses historical data and simplified equations to estimate the mission performance and the weights breakdown of an aircraft. Because it is mainly based on simpler equations, its run time for a single case is very low, making it possible to run a relatively high number of cases. Because FLOPS uses lumped parameters, it is only logical to go to a slightly higher fidelity tool that is appropriate with the conceptual design phase in order to break down the lumps. Therefore, a more detailed analysis of the epistemic uncertainty variables will be possible.

OpenAeroStruct~\cite{Jasa2018} is a lightweight, open-source tool designed for integrated aerostructural analysis and optimization. It combines aerodynamic and structural analysis capabilities within a gradient-based optimization framework, enabling efficient design of aircraft structures. The tool supports various analyses, including wing deformation effects on aerodynamics and the optimization of wing shape and structure to meet specific design objectives. In this work it will be used as a step following FLOPS anaylses, at it represents a step increase in fidelity. According to the selected analysis tools, the main parameter uncertainties involved are to be investigated are the material properties (e.g., Young's modulus) and the aerodynamic properties (e.g., lift and drag coefficients).

\begin{table}[]
\centering
\caption{Nomenclature for mentioned FLOPS variables}
\begin{tabular}{ll}
\multicolumn{1}{c}{\textbf{Variable Name}} & \multicolumn{1}{c}{\textbf{Description}}                                               \\ \hline
WENG                   & Engine weight scaling parameter                                                                             \\
OWFACT                 & Operational empty weight scaling parameter                                                                  \\
FACT                   & Fuel flow scaling factor                                                                                    \\
RSPSOB                 & Rear spar percent chord for BWB fuselage at side of the body                                                \\
RSPCHD                 & Rear spar percent chord for BWB at fuselage centerline                                                      \\
FCDI                   & \begin{tabular}[c]{@{}l@{}}Factor to increase or decrease lift-dependent drag\\ coefficients\end{tabular}   \\
FCDO                   & \begin{tabular}[c]{@{}l@{}}Factor to increase or decrease lift-independent drag\\ coefficients\end{tabular} \\
FRFU                   & Fuselage weight (composite for BWB)                                                                         \\
E                      & Span efficiency factor for wing                                                                             \\
\end{tabular}
\label{tab:flops_nomenclature}
\end{table}

First, among the list of FLOPS input parameters, 31 of them are selected as they are either not related to the design, or highly abstracted parameters that may capture the highest amount of abstraction and cause variance in the outputs. Of these 31 parameters, 27  of them are scaling factors and are assigned a range between 0.95 and 1.05. Remaining four are related to the design, such as spar percent chord at BWB at fuselage and side of the body and they are swept in estimated, reasonable ranges. The parameter names and their descriptions will be explained throughout the discussion of the results as necessary, but an overview of them are listed in Table~\ref{tab:flops_nomenclature} for the convenience of the reader. Aircraft range is calculated through a combination of sampled input parameters. Among these parameters 4096 samples are generated using Saltelli sampling~\cite{Saltelli2002}, a variation of fractional Latin Hypercube sampling, due to its relative ease in calculation of Sobol indices and availability of existing tools. Calculation of the indices are carried out by using the Python library SALib~\cite{Iwanaga2022}.

\begin{figure}[]
	\centering
	\includegraphics[width=0.66\linewidth]{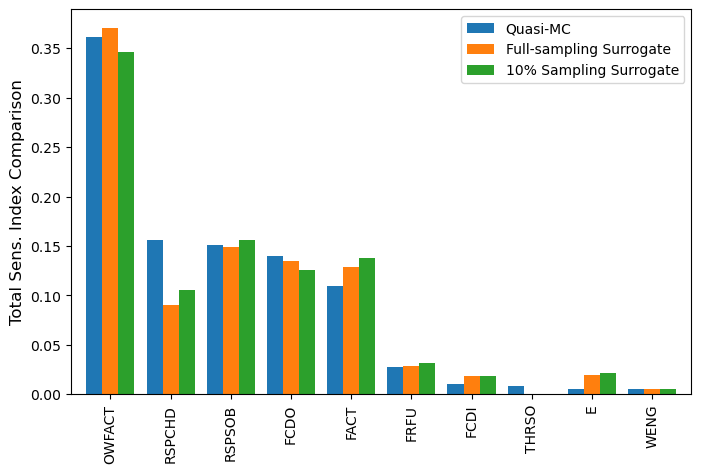}
	\caption{\centering Comparison of sensitivity indices calculated by three different methods: Quasi Monte Carlo, Surrogate model with full sampling, and surrogate model with 10\% sampling}
	\label{fig:ST_comp_three}
\end{figure}

In Figure~\ref{fig:ST_comp_three}, total sensitivity indices calculated are given for three different sampling strategies. Blue bars represent the Quasi-Monte Carlo (QMC) sampling that uses all 4096 input factors. To demonstrate the impact of how sensitivity rankings may change through the use of surrogate modeling techniques, two different response equations (RSE) are employed. First one is the an RSE that is constructed using all 4096 points, and the other one is constructed using only 10\% of the points, representing a degree of  increase in the computational cost. After verifying that these models fit well, it is seen that they are indeed able to capture the trends albeit the ranking of important sensitivities need to be paid attention. 

In this analysis, it is seen that aerodynamic properties, material properties and the wing structure location have been found to have significant impact on the wing weight and aerodynamic efficiency. Therefore, they are identified as critical uncertainties. This is indeed expected and consistent with the results found for a tube-and-wing configuration before in Reference~\cite{Yarbasi2022}.


\subsection{ Experiment Design and Execution}

For demonstration of the methodology, the next step will include the low-fidelity aero-structures analysis tool, OpenAeroStruct~\cite{Jasa2018} and how such a tool can be utilized to guide the design of a physical experimentation setup.

    \begin{enumerate}
        \item \textit{Problem Investigation}:
        \begin{itemize}
            \item \textit{Premise}: The variations in range calculations are significantly influenced by uncertainties in the Young's modulus, wingbox location and aerodynamic properties. These parameters are related to the disciplines of aerodynamics and structures.
            \item \textit{Research Question}: How can the uncertainties in these parameters impacting the range can be reduced?
            \item \textit{Hypothesis}: A good first-order approximation is the Breguet range equation is used to calculate the maximum range of an aircraft~\cite{Cavcar2006}. A more accurate determination of the probability density function describing the aerodynamic performance of the wing will reduce the uncertainty in the wing weight predictions.
        \end{itemize}
        \item \textit{Thought Experiment}: Visualizing the impact of more accurately determined parameters on the simulation results, we would expect to see a reduction in the variation of the simulation outputs.
        \item \textit{Purpose of the Experiment}: This experiment aims to reduce the parameter uncertainty in our wing aero-structures model. There is little to none expected impact of unknown physics that would interfere with the simulation results at such a high level of abstraction. In other words, the phenomenological uncertainty is expected to be insignificant for this problem. 
        In order to demonstrate the proposed methodology, both avenues ---computational experiment only, and physical experiment--- will be pursued.
        \item \textit{Experiment Design}: We decide to conduct a computational experiment that represents a physical experimentation setup where the parameters pertaining to the airflow, material properties and structural location are varied within their respective uncertainty bounds and observe the resulting lift-to-drag ratios. For subscale physical experiment, the boundary conditions of the experiment need to be optimized for the reduced-scale so the closest objective metrics can be obtained. 
        \item \textit{Computational Experiments for both cases}:
        \begin{itemize}
            \item \textit{Define the Model}: We use the OpenAeroStruct wing structure model with a tubular spar structures approximation and a wingbox  model.
            \item \textit{Set Parameters}: The parameters to be varied are angle of attack, Mach number, location of the structures.
            \item \textit{Design the Experiment}: We use a Latin Hypercube Sampling to randomly sample the parameter space. Then Sobol indices are computed to observe the global sensitivities over the input space.
            \item \textit{Develop the Procedure}: 
            \begin{itemize}
                \item \underline{For CX only:} For each random set of parameters, run the OpenAeroStruct model and record the resulting predictions, case numbers and run times. After enough runs to sufficiently explore the parameter space,  analyze the results.
                \item \underline{For PX only:} Use the wingbox model only in OpenAeroStruct, pose the problem as a constrained optimization problem to get PX experimentation conditions, scale is now a design variable, scale dimensionless parameters accordingly.
                
            \end{itemize}
        \end{itemize}
    \end{enumerate}

\subsection{Computer Experiments for Uncertainty Mitigation}

Reducing the uncertainty in the lift-to-drag ratio would have a direct impact on reducing the uncertainty in range predictions. L/D is a key aerodynamic parameter that determines the efficiency of an aircraft or vehicle in converting lift into forward motion while overcoming drag. By reducing the uncertainty in L/D, one can achieve more accurate and consistent estimates of the aircraft's efficiency, resulting in improved range predictions with reduced variability and increased confidence.


To calculate L/D and other required metrics, the low-fidelity, open-source aerostructures analysis software OpenAeroStruct is used. BWB concept illustrated in Figure~\ref{fig:BWB_VSP} is exported to OpenAeroStruct. For simplicity purposes, vertical stabilizers are ignored in the aerodynamics and structures analyses. The wingbox is abstracted as a continuous structure, spaning from 10\% to 60\% of the chord, throughout the whole structural grid. This setup is used for both CX and PX cases, and the model parameters are manipulated according to the problem.

\subsubsection{Test conditions}

First, it is necessary to develop a simulation that would develop the full-scale conditions. The simpler approximation models the wingbox structure as a tubular spar, with a reducing diameter from root to tip. The diamaters are calculated through optimization loops so that stress constraints are met. For the wingbox model, the location is approximated from the conceptual design of the structrual elements from public domain knowledge. For both cases, different aerodynamic and structural grids are employed to investigate the variance in SRQs. Cruise conditions at 10000 meters are investigated, with a constant Mach number of 0.84.

\begin{figure}
    \centering
    \includegraphics[width=0.7\linewidth]{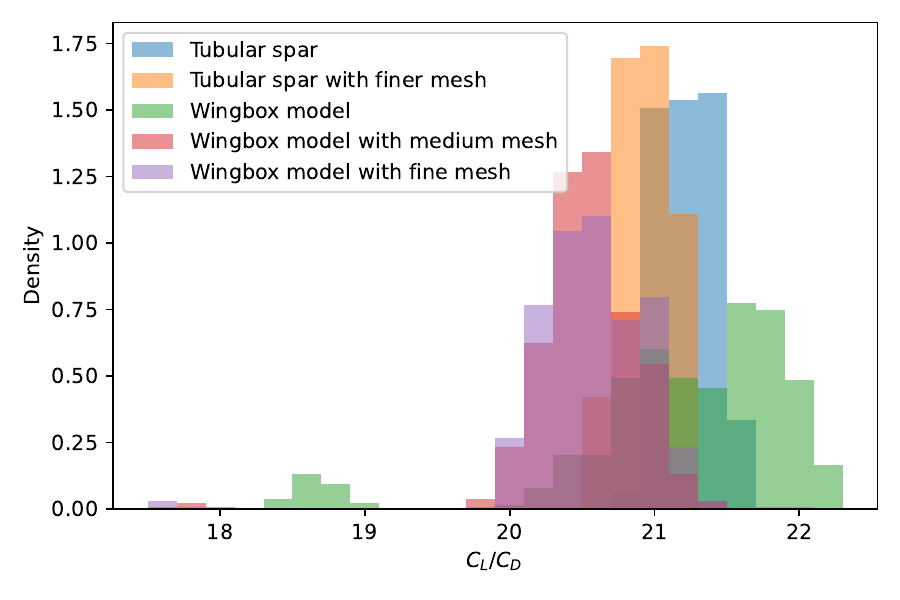}
    \caption{Probability densities of $C_L / C_D$ for five different model structures.}
    \label{fig:clcd_densities}
\end{figure}

Five different model structures are tested for this experimentation setup with the same set of angle of attack, Mach number, spar location and Young's modulus in order to make an accurate comparisons.  Through these runs, lift-to-drag ratios are calculated and the histogram is plotted in Figure~\ref{fig:clcd_densities}. In this figure, the first observation is that although the wingbox model is a better representation of reality, its variance is higher than compared to the tubular spar models, as well as being hypersensitive to certain inputs in some conditions.  It is also seen that the predictions of the tubular-spar model generally lie between the predictions that of the two different fidelities of the wingbox model. 

Furthermore, the runtime statistics have a significant impact on how the results are interpreted, as well as how many cases can be realistically considered. The overview is presented in Table~\ref{tab:run_comparison}. It is obvious that the mesh size is the most dominant factor on the average run time for a single case.  An interesting observation is that the coarser wingbox model takes less time to run compared to a lower fidelity tubular spar model, and predicts a higher lift-to-drag ratio. The reason of this was the wingbox model with the coarser mesh was able to converge in fewer iterations compared to the tubular-spar model. Using a shared set of geometry definition and parameters, as much as the corresponding fidelity level allows, showed that decreasing the mesh size resulted in less variance in the predicted SRQs, as expected. However, increasing the fidelity level comes with a new set of assumptions pertaining to the newly included  subsystems, or physical behavior. Therefore, one cannot definitively say that increasing the fidelity level would decrease the parameter uncertainty without including the impact of the newly added parameters. 


\begin{table}[htbp]
    \centering
    \caption{\centering OpenAeroStruct runtime and output variation statistics with respect to different model structures, on a 12-core 3.8 GHz 32GB RAM machine}
    \begin{tabular}{ccc}
        \hline
        Run type & Std. Deviation in L/D & Mean Runtime [s]\\
        \hline
        Tubular spar-coarse   & 0.184     & 6.79 \\
        Tubular spar-medium   & 0.169     & 15.38 \\
        Wingbox - coarse  & 0.794    & 2.5 \\
        Wingbox - medium & 0.376     & 20.7 \\
        Wingbox - fine & 0.401 & 76.67 \\
        \hline
    \end{tabular}
    \label{tab:run_comparison}
\end{table}

\subsection{Leveraging Computer Experiments for Guiding Physical Experimentation}

\subsubsection{Feasibility of the full-scale model}

As discussed before, construction and testing of a full-scale vehicle models is almost always not viable in the aerospace industry, especially in the earlier design phases. For this demonstration, a free-flying sub-scale test will be pursued. The baseline experiment conditions will be the same as in the computational-only experimentation, except for appropriate scaling of parameters. Therefore, a scale that would  be an optimum for a selected cost function needs to be found, considering the constraints. For this use case, following constraints are defined:

\begin{itemize}
    \item Scale of the sub-scale model: $n<0.2$
    \item Mach number: $0.8<Ma<0.87$, The effects of compressibility are going to be much more dominant as $Ma=1$ is approached, therefore the upper limit for the Mach number was kept at 0.87.
    \item Angle of attack $0<\alpha<10$, Because all similitude conditions will not be met, it the flight conditions for a different angle of attack need to be simulated. This is normal practice in subscale testing~\cite{Wolowicz1979}.
    \item Young's Modulus:  $E_S< 3 E_F$,Young's modulus of the model, should be less than three times of the full-scale design.
\end{itemize}

\begin{figure}
    \centering
    \includegraphics[width=0.75\linewidth]{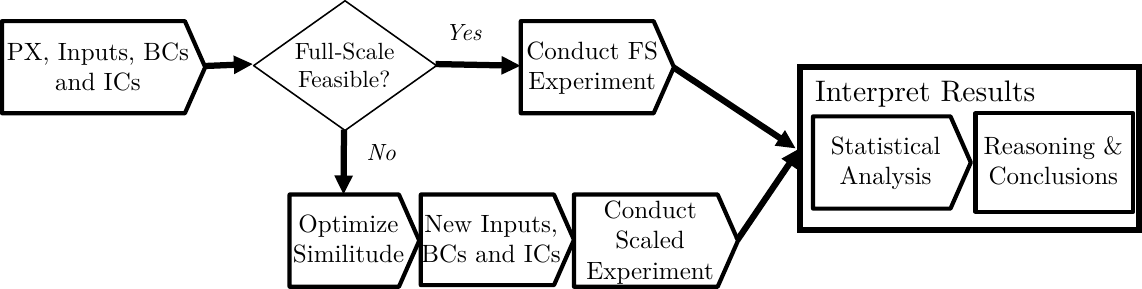}
    \caption{Process for sub-scale experiment design. Note that this is a derivative of the process given in Figure~\ref{fig:rq3_overview}}
    \label{fig:scaled_experiment}
\end{figure}

\subsubsection{Optimize for similitude}

For this optimization problem, a Sequential Least Squares Programming method is used. SLSQP is a numerical optimization algorithm that is particularly suited for problems with constraints~\cite{Lyu2014}. It falls under the category of sequential quadratic programming (SQP) methods, which are iterative methods used for nonlinear optimization problems. The idea behind SQP methods, including SLSQP, is to approximate the nonlinear objective function using a quadratic function and solve a sequence of quadratic optimization problems, hence the term ``\textit{sequential}". In each iteration of the SLSQP algorithm, a quadratic sub-problem is solved to find a search direction. Then, a line search is conducted along this direction to determine the step length. These steps are repeated until convergence. One advantage of SLSQP is that it supports both equality and inequality constraints, which makes it quite versatile in handling different types of problems. It is also efficient in terms of computational resources, which makes it a popular choice for a wide range of applications.

\begin{algorithm}
\caption{Constrained Optimization for finding Physical Experiment Conditions}
\begin{algorithmic}
\Procedure{Optimization}{x}
\State Define Scaling parameters
\State Define x=[$n$, $\alpha$, Ma, $h$, E]
\State Define constraint s
\State Initialize x with initial guess [0.1, 0, 0.84, 10000, 73.1e9]
\While{not converged}
\State Evaluate cost function f(x)  \Comment{Equation~\ref{eq:cost_function}}
\State Solve for gradients and search directions
\State Run OpenAeroStruct optimization 
\If{Failed Case}
\State Return high cost function
\State Select new x
\EndIf
\EndWhile
\State \Return x
\EndProcedure
\end{algorithmic}
\label{alg:rq3.3}
\end{algorithm}

The algorithm used for this experiment is presented in Algorithm~\ref{alg:rq3.3}. For convenience purposes, the altitude is taken as a proxy for air density. In the optimization process, mass (including the fuel weight and distribution) is scaled according to:
\begin{equation}
    n_{mass} = \frac{\rho_F}{\rho_S} n^3
\label{eq:mass_scaling}
\end{equation}
where $\rho_F$ is the fluid density for the full-scale model, $\rho_S$ the fluid density for the sub-scale model, and $n$ is the geometric scaling factor~\cite{Wolowicz1979}. Since aeroelastic bending and torsion are also of interest, following aeroelastic parameters for bending ($S_b$) and torsion ($S_t$) need also be satisfied, and they are defined as:

\begin{equation}
    S_b  = \frac{EI}{\rho V^2 L^4}
    \label{eq:bending_similitude}
\end{equation}

\begin{equation}
    S_t  = \frac{GJ}{\rho V^2 L^4}
    \label{eq:torsion_similitude}
\end{equation}

These two parameters need to be duplicated in order to assure the similitude of inertial and aerodynamic load distributions for the same Mach number or scaled velocity, depending on the compressibility effects at the desired test regime~\cite{Wolowicz1979}. And the cost function is selected to be:

\begin{equation}
    f(x) =  \left|\frac{\frac{C_L}{C_D}_S - \frac{C_L}{C_D}_F}{\frac{C_L}{C_D}_F} \right|^2 + 30 \left|\frac{Re_S - Re_F}{Re_F} \right|^2 + 3000 \left|\frac{Ma_S - Ma_F}{Ma_F}\right|^2
    \label{eq:cost_function}
\end{equation}

where, lift-to-drag ratio, Reynolds number and the Mach number of the sub-scale model are quadrically penalized with respect to their corresponding deviation form the simulation results of the full scale result. Because the magnitudes of the terms are vastly different, second and third terms are multiplied with coefficients that would scaled their impact to the same level of the first term. For other problems, these coefficients present flexibility for engineers. Depending on how much certain deviations in the ratio of similarity parameters are penalized, the optimum scale and experiment conditions will change. In this application, the simulated altitude for the free-flying model is changed, rather than changing the air density directly.

\subsubsection{Interpret results}

Without the loss of generality, it can be said that the optimum solution may be unconstrained or constrained, depending on the nature and the boundaries of the constraints. At this step, the solution will need to be verified as to whether it will lead to a feasible physical experiment design. Probable causes will be conflicts between the design variables, or infeasibility to match them in a real experimentation environment. In such cases, the optimization process needs to be repeated with these constraints in mind. However, for this problem, the constrained optimum point is found to be:

\begin{itemize}
    \item $n = 0.2 $
    \item $Ma = 0.86$
    \item $\alpha =10$
    \item $E_S = 219GPa$
    \item $h = 0m$
    \item $Re = 6.2x10^6$   
\end{itemize}

Furthermore,, it can be noted that due to the nature of the constrained optimization problem, the altitude for the sub-scale test was found to be sea-level. While, this is not completely realistic, it points to an experiment condition where high-altitude flight is not necessary. Finally, the Young's modulus for the sub-scale model is slightly below the upper threshold, which was three times that of the Young's modulus of the full-scale design. With this solution we can verify that solving a constrained optimization problem to find the experiment conditions is a valid approach, and provides a baseline for other sub-scale problems as well.

\begin{sidewaysfigure}
    \centering
    \includegraphics[width=0.9\linewidth]{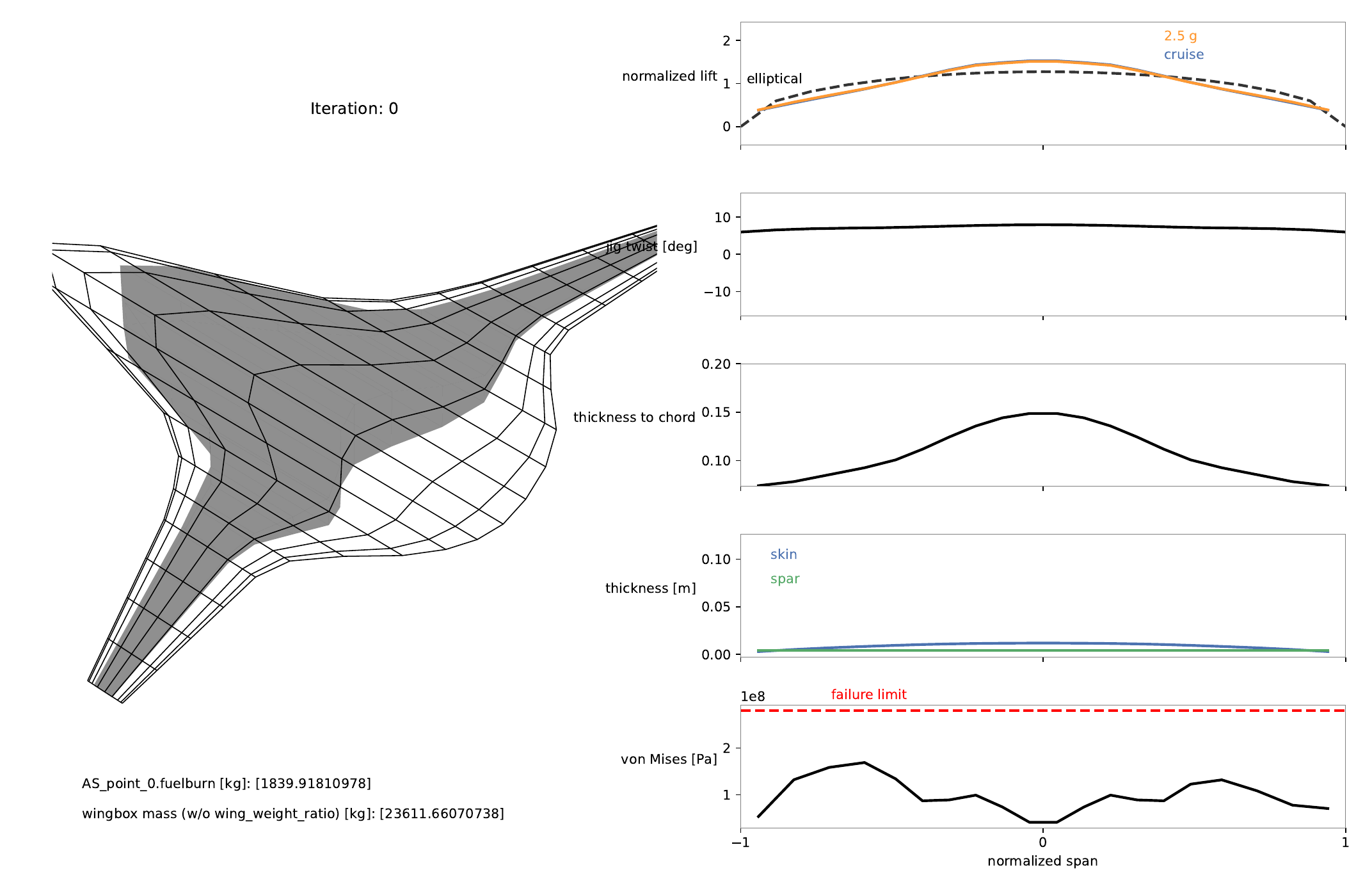}
    \caption{BWB vortex-lattice mesh with the structural wingbox model in grey shade. On the right hand side, results of an example run are shown.}
    \label{fig:bwb_aero_conv}
\end{sidewaysfigure}

\section{Conclusion}

In this work, a novel approach is introduced for the design of physical experiments, emphasizing the quantification of uncertainty to target the of engineering models with a specific focus on early-stage aircraft design. Sensitivity analysis techniques are intelligently utilized to find out specific computational and physical experimentation conditions, to tackle the challenge mitigation of epistemic uncertainty. Findings indicate that this methodology not only facilitates the identification and reduction of critical uncertainties through targeted experimentation but also optimizes the design of physical experiments through computational efforts. This synergy enables more precise predictions and efficient resource utilization. Through a case study on a Blended-Wing-Body (BWB) aircraft concept, the practical application and advantages of the proposed framework are exemplified, demonstrating how subsequent fidelty levels can be leveraged for uncertainty mitigation purposes.

Presented framework for uncertainty management that is adaptable to various design challenges. The study highlights the importance of integrating computational models by guiding physical testing, fostering a more iterative and informed design process that will save resources. Of course, every problem and testing environment got its own challenges. Therefore, dialogue between all parties involved in model development and physical testing is encouraged.

Future research is suggested to extend the application of this methodology to different aerospace design problems, including propulsion systems and structural components. Additionally, the development of more advanced computational tools and algorithms could further refine uncertainty quantification techniques. With more detailed models and physics, integration of high performance computers, it is possible to see the impact of this methodology in later stages of the design cycle. The reduction of uncertainty on performance metrics can contribute to avoiding program risk --- excessive cost, performance shortcomings and delays.

\section*{Acknowledgments}

This work represents a continuation of the efforts undertaken by the NATO Applied Vehicle Technologies-297 team. The author gratefully acknowledges valuable discussions with Drs. Nigel Taylor, Burak Bağdatlı, and Eric Walker, who contributed to shaping the methodology presented in this paper. Supplementary files regarding the BWB optimization can be found in \url{https://github.com/eyyarbasi/BWB-opt-supp}.


\bibliography{sample}

\end{document}